# Photonic Floquet topological insulators in a fractal lattice

Zhaoju Yang, Eran Lustig, Yaakov Lumer and Mordechai Segev†

*Physics Department and Solid State Institute,*
*Technion– Israel Institute of Technology, Haifa 32000, Israel*
†Corresponding author. Email: msegev@technion.ac.il

## Abstract

**We present Floquet fractal topological insulators: photonic topological insulators in a fractal-dimensional lattice consisting of helical waveguides. The helical modulation induces an artificial gauge field and leads to the trivial-to-topological transition. The quasienergy spectrum shows the existence of topological edge states corresponding to real-space Chern number 1. We study the propagation of light along the outer edges of the fractal lattice and find that wavepackets move along the edges without penetrating into the bulk or backscattering even in the presence of disorder. In a similar vein, we find that the inner edges of the fractal lattice also exhibit robust transport when the fractal is of sufficiently high generation. Finally, we find topological edge-states that span the circumference of a hybrid half-fractal half-honeycomb lattice, passing from the edge of the honeycomb lattice to the edge of the fractal structure virtually without scattering, despite the transition from two-dimensions to a fractal dimension. Our system offers a realizable experimental platform to study topological fractals and provides new directions for exploring topological physics.**

**Introduction**

Topological insulators are a new phase of matter, characterized by the insulating bulk and conductive edges [1,2]. They have been at the forefront of the condensed matter physics for the past decade and more recently inspired the emergence of topological phases in many classical-wave systems [3-5], such as microwaves [6-8], photonics [3, 6-15], acoustics [4] and more. Photonics specifically has become the cutting edge platform to explore all kinds of topological phases ranging from the quantum Spin Hall effect [11], topological crystalline insulator [15] and valley Hall effect [16,17] all the way to topological systems that lack periodicity, such as topological quasicrystals [18] and even topological Anderson insulators in which topology is induced by disorder [19]. However, thus far all studies of topological insulators have explored systems in integer dimensions (physically, 2D or 3D) with well-defined bulk and edges. But the physical dimensions do not always define the dimensions in which a system evolves: some structures have a non-integer (fractal) dimension, despite being in a 2D or 3D realm. The existence of systems with fractal dimensions raises a series of fascinating questions in the context of topological physics. For example, is it possible to realize topological edge states in fractal dimensions? Moreover, fractal structures tend to include holes, so can topological edge states be found around every hole in the system or only in the external boundary? Intuitively, one might think that there are no topological edge states because of the bulk-edge correspondence [20,21] and the lack of a true bulk. This raises a deeper question: is there bulk-edge correspondence when the fractal structure is actually made up of holes within the bulk?

Here, we investigate the photonic Floquet topological phase in a periodically driven fractal lattice. This lattice relies on the fractal photonic crystal [Sierpinski gasket (SG)] consisting of evanescently coupled helical waveguides, which can be realized by the femtosecond-laser-writing technology [22]. We calculate the topological Floquet spectrum and show the existence of topological edge states corresponding to real-space Chern number 1 [23,24], which can be controlled by the periodic driving. We explore the dynamics of the edge states and their robustness in simulations in the fractal SG lattice, and find that wavepackets made up of topological edge states propagate along the outer edge without penetration into the bulk and without backscattering even in the presence of disorder and sharp corners. Likewise, the topological edge states associated with inner edges in the fractal lattice exhibit robust transport whenever the inner edge includes a large enough area (greater than the third generation of the SG fractal). These results imply that fractal structures can act as topological insulators, despite the lack of periodicity and being made up mostly of holes. Finally, we study transport in a hybrid lattice combining the fractal lattice with a honeycomb lattice, and find that topological edge states can pass from the honeycomb lattice into the edge of the fractal lattice, where they exhibit topologically-protected transport. This further proves that the edge states in the fractal lattice directly correspond to the same Chern number as the honeycomb lattice driven by the same periodic modulation. Finally, it is possible to obtain similar results with other fractal platforms: the Sierpinski carpet under aperiodic arrangement similar to Ref. [13], and conjecture that the 3D realizations of both the Sierpinski gasket and the carpet also give rise to topological edge states, and likewise the Cantor cubes and Cantor dust. Hence, our results suggest a wealth of new kinds of topological systems and new applications, such as using the topological robustness combined with the enhanced sensitivity of fractal systems for sensing, and, in non-Hermitian settings - topological insulator lasers [25,26], in fractal dimensions.

**Results**

Our starting point is the Sierpinski gasket with a Hausdorff dimension $d_f = \ln(3)/\ln(2) = 1.585$. Consider a photonic lattice of evanescently-coupled helical waveguides, similar to Ref. [12]. Figure 1 shows the iterative generations of the SG. As can be seen, the first generation G(1) of SG has 9 blue circles, which indicate the positions of the helical waveguides. The generation G(2) consists of three copies of G(1), sharing three vertices. Accordingly, the G(2) waveguide lattice has 24 waveguides organized as the second generation of the SG. Similarly, G(n) has three copies of G(n-1), sharing three corner sites. Hereafter, we focus on the fractal lattice of generations G(4) and G(5), and we conjecture that the conclusions we draw from this study hold for the SG lattice in any dimension. Examination of Fig. 1 reveals that all sites in the SG fractal lattice are on the boundaries, and there is not even a single site that does not reside on a boundary – external or internal. Finally, as in Ref. [12], this lattice consists of helical waveguides, which is equivalent to a periodically driven potential that introduces an artificial gauge field $\mathbf{A}$.

The equation governing the diffraction of light in this fractal photonic lattice under the tight-binding approximation [12] can be written as

$$i\partial_z\psi_n = c_0 \sum_{\langle m \rangle} e^{i\mathbf{A}(z)\cdot \mathbf{r}_{m,n}}\psi_m \qquad (1)$$

where $z$ is the distance of the light propagation along the waveguide, $\psi_n$ is the amplitude of the electric field in $n$th waveguide, $c_0$ is the coupling strength between two nearest sites, $\mathbf{r}_{m,n}$ is the displacement vector pointing from waveguide $m$ to waveguide $n$, $\mathbf{A}(z) = A_0[\sin(\Omega z), -\cos(\Omega z), 0]$ is the artificial vector potential induced by the helicity of the waveguides with amplitude $A_0 = kR\Omega$, where $k$ is the wavenumber of the light in the medium, $R$ is the radius of the helix, $\Omega$ is the longitudinal frequency of the helix corresponding to periodicity $L = 2\pi/\Omega$, and $\langle m \rangle$ indicates that the summation is taken over all the nearest waveguides to waveguide $n$. The light evolution in the system is described by the paraxial wave equation, which is mathematically equivalent to the Schrödinger equation with $z$ axis playing the role of time. Equation (1) is derived by applying the tight-binding approximation to the paraxial wave equation.

The eigen-values and eigen-states can be obtained by diagonalizing the unitary evolution operator for one period [27]. The results of the quasi-energy spectrum $\beta$ (which in a photonic lattice are the deviation of the propagation constant from the wavenumber in the medium, [12]) in the fractal SG systems are shown in Fig. 2(a), with $\mathbf{A}(z) = 0$ corresponding to the straight waveguides and Fig. 2(b) with $\mathbf{A}(z) \neq 0$ corresponding to the helical waveguides. The spectrum for our G(4) fractal lattice is organized in 5 bunches ('bands') separated by gaps (grey shaded regions in Fig. 2(a,b)). The spectrum of the non-topological system (Fig. 2a) shows a large central gap, with a flat "band" in mid-gap. These states are immobile and degenerate (all have the same energy), as expected from a non-topological system. On the other hand, for the driven system (the helical waveguides), as shown in the Fig. 2b, the edge states from the central flat-band evolved into non-degenerate unidirectional edge states. Figure 2(c) shows the field intensities of the actual wavefunctions of such eigen-states, specifically states number 93 and 95, with quasi-energies -0.040 and -0.018, respectively. These states are localized at the exterior (state number 95) and the interior edges (state number 93). As we see next, these states behave as topological edge states, exhibiting Chern number 1 and topologically-protected transport. Note that in other bunches with quasienergies below -0.2 or above 0.2, the eigenstates are bulk states.

To verify that the edge states we have found [the non-degenerate unidirectional states in the rectangle of Fig. 2(b), two of which are shown in Fig. 2(c)] are indeed topological, we need to characterize our system through its Chern number. Since fractal lattices here are non-periodic, we calculate the real-space Chern number [23,24]. Heuristically, the real-space Chern number "measures" the chirality of states at a specific quasi-energy, and in periodic systems yields the same integer number as the "standard" Chern number (defined on momentum space) [23,24]. The definition of the real-space Chern number is:

$$C = 12\pi i \sum_{j \in A} \sum_{k \in B} \sum_{l \in C} (P_{jk} P_{kl} P_{lj} - P_{jl} P_{lk} P_{kj}), \quad (2)$$

where $j, k, l$ are the lattice sites indices within three different neighboring regions A, B, C [as drawn in the inset of Fig. 3(b), arranged counter-clockwise], $P_{jk} = \langle j|P|k \rangle$ and the projector operator $P$ projects onto a given state of a specific quasienergy (a state with quasienergy playing the role of the Fermi level). The results are shown in Fig. 3(a, b). We calculate the real-space Chern number for our fractal lattice, and, for a direct comparison, also for a honeycomb lattice, when both are driven by the same periodic modulation (manifested here as the helicity of the waveguides). The lower panels in Fig. 3 show the methodology of the calculation: the hexagons are divided into three distinct regions (A,B,C), each enclosing many helical waveguides, for both the fractal and the honeycomb lattices. As expected, the helicity induces a topological bandgap in the honeycomb lattice [Fig. 3(a)] corresponding to real-space Chern number 1, which coincides with the outcome of the natural momentum-space calculation of the Chern number (that can be used here because the honeycomb lattice is periodic). For the fractal lattice [Fig. 3(b)], the result of the real-space calculation is interesting, as there are many quasienergy values having nonzero real-space Chern number. The most important quasi-energy range is from -0.05 to 0.05, which is within the topological bandgap of the helical honeycomb lattice where the real-space Chern number is 1. As shown in Fig. 3(b), in the helical fractal lattice –the quasienergies in the range between -0.05 to 0.05 correspond to Chern number 1, hence support the observation that edge states in this range (e.g., state number 95 and 93) are indeed topological. We find that the gaps between regions of eigen-values around -0.5, -0.2, 0.2 and 0.5, marked by grey shading in Fig. 2(b), separate different bunches of "bulk states" (states residing away from the edges) with real-space Chern number of 0, which means that these gaps are topologically trivial.

Having found unidirectional edge states with real-space Chern number 1, we study the evolution of the edge states in evolution simulations, in the presence of defects and disorder. Specifically, to verify the edge state number 95 is indeed topological, we demonstrate its ability to display topologically protected transport, the hallmark of topological physics. We launch a wavepacket at the edge of the fractal lattice, and simulate its propagation (Fig. 4(a-e)). The initial wavepacket is a superposition of eigen edge states such that it has a finite width (see Fig 4a). Figures 4(b-e) show the light intensity at different propagation distances $Z = 10, 20, 30, 40$cm. Clearly, the wavepacket moves along the edge of the fractal lattice, passes the corner without scattering. During the propagation, the wavepacket remains confined to the edge, not penetrating into the bulk and backscattering. Next, we test the robustness to disorder. The simulation in Fig. 4(f-j) shows that the wavepacket can pass a defect (indicated by the blue dot in the fractal lattice) – a site with on-site disorder of strength $0.1c_0$. We find that the propagation of such wavepackets of edge states in the fractal system is very robust against random on-site disorder of strength up to $0.2c_0$. The only visible difference between the initial and final waveapckets is the diffraction-broadening, caused by dispersion (because the edge states comprising the wavepacket evolve at slightly different rates).

The topologically protected transport of edge states in the fractal lattice is not unique to the outer edge. The Supplementary Movie #1 show a similar simulation for an inner edge in the fractal lattice (the perimeter of a hole). The excited edge state number 93 exhibits robust evolution, in the same vein as for the outer edge of the fractal lattice. Since higher generation fractals always include more and more inner edges as the generation increases, we find (in simulations) that they exhibit robust propagation on the inner edges - as long as the edge includes an area that is larger than G(3), so as to serve as the "bulk" region for the respective inner edge.

Altogether, we have shown that the fractal lattice of helical waveguides has non-degenerate unidirectional edge state residing in a gap (Fig. 2), that several edge states have real-space Chern number of 1 (Fig. 3), and that wavepackets made up of these edge states (in both outer and inner edges) display robust transport by going around corner and passing defects without backscattering or scattering into the bulk. ***Hence, we proved that the fractal lattice acts as a topological insulator, despite the fact that there is no bulk whatsoever, and that we cannot rely on bulk-edge correspondence.***

At this point it is very important to emphasize that there are key differences between the fractal lattice and a helical honeycomb lattice with randomly missed sites. As we show in the Supplementary Material, section B, this honeycomb lattice with randomly missed sites is not a topological insulator: its real-space Chern number is always in the proximity of zero, and its 'edge states' do not exhibit unidirectional robust transport. It is clear that the additional symmetries of self-similarity on multiple scales, which is at the heart of fractality, is crucial for the existence of topological features in driven fractal lattices.

Finally, we study a hybrid lattice combining both the fractal and the honeycomb lattices, stitched together, as shown in Fig. 5. We launch a wavepecket comprising of topological edge states on the honeycomb side, and simulate its propagation into the fractal side of the lattice. Had our lattice been strictly honeycomb – such a wavepacket would propagate without scattering into the bulk and without backscattering even in the presence of disorder (or defects) - as long as the amplitude of the disorder (defect) does not close the topological gap. But our lattice here is a hybrid: half honeycomb half fractal. Hence, this numerical experiment will serve to show whether (or not) the edge states we have found support topologically-protected transfer from honeycomb to fractal lattices, modulated by the same helicity. The launched wavepacket, shown in Fig. 5(a), is constructed from a superposition of edge states of the honeycomb lattice. Figures 5(b-d) show the evolution, displaying the light intensity distributions at several propagation distances $Z = 5, 10, 15$cm. The wavepacket moves along the edge of the honeycomb lattice, passes the corner without scattering, enters the fractal lattice and continues moving along the edge of the fractal lattice. Throughout propagation in the hybrid lattice, the wavepacket remains confined to the edge, not penetrating into the bulk and without any backscattering. Moreover, the simulation in Fig. 5(e-h) shows that the wavepacket is able to pass a defect in the fractal lattice (its position is given by the blue dot) – a site with on-site disorder of strength $0.1c_0$. Supplementary Movies #2-3 show long term propagation in this hybrid lattice, with the wavepacket encircling the lattice multiple times. Supplementary Movie #4 shows the transport with the wavepacket initially launched at the fractal lattice. Finally, Fig. S3 shows the propagation of a wavepacket in a hybrid lattice where the two components possess different non-zero real space Chern number. In such a non-matched semi-fractal lattice the wave partially moves along the edge and partially penetrates into the 'bulk' of the fractal lattice, which indicates that such a system has no topological protection. That is, for a hybrid semi-fractal system to be topological – its constituents should have the same real-space Chern number.

**Discussion**

As stated earlier, our findings here are in fact a prelude to upcoming experiments in a photonic platform, which will provide experimental proof that indeed fractal lattices can be made topological. It is therefore essential to carry out wave dynamics simulations with the actual experimental parameters, and examine the evolution. As shown in the Supplementary Material Fig. S4, we simulate the wave dynamics of the tight-binding example of Fig. 5. Our wave simulations show good agreement with the tight-binding simulations, suggesting that indeed what we propose here is readily experimentally accessible with current technology.

In summary, we proposed photonic Floquet topological insulators in a fractal lattice, and demonstrated robust transport along the outer and inner edges of the fractal landscape. We underpinned the difference between a driven (helical) fractal lattices and lattices with randomly missed sites, and showed that the fractal symmetries are crucial for the existence of the topological features. Finally, we showed that topological edge states can pass from the edge of the honeycomb lattice to the outer edge of the fractal structure (of the same chirality) virtually without scattering, despite the transition from two-dimensions to a fractal dimension. The parameters used in this work are all readily accessible in experiments with photonic lattices fabricated using direct laser writing [12,

22]. Such experiments could be the first experimental realization of topological fractal insulators [28].

**Acknowledgements**

This work was sponsored by the Israel Science Foundation and by an Advanced Grant from the European Research Council.

**Conflict of interest**

The authors declare no conflicts of interest.

**Author contributions**

All authors contributed significantly to this work.

**Supplementary information** is available online.

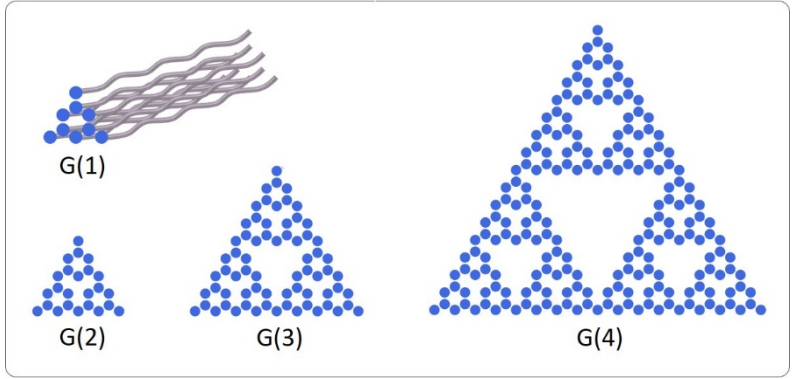

Figure 1.

Iterative generations of the Sierpinski gasket (SG). The first generation G(1) of the SG has 9 blue sites. G(n) has three copies of G(n-1), sharing three corner sites. The fractal lattice of helical waveguides is the generation G(4) with in total 204 sites. Each blue site marks the position of a helical waveguide. The helicity of the waveguides introduces an artificial vector potential $\bm{A}(z) = A_0[\sin(\Omega z), -\cos(\Omega z), 0]$. For presentation simplicity, we draw only the three-dimensional schematic of a G(1) lattice of helical waveguides.

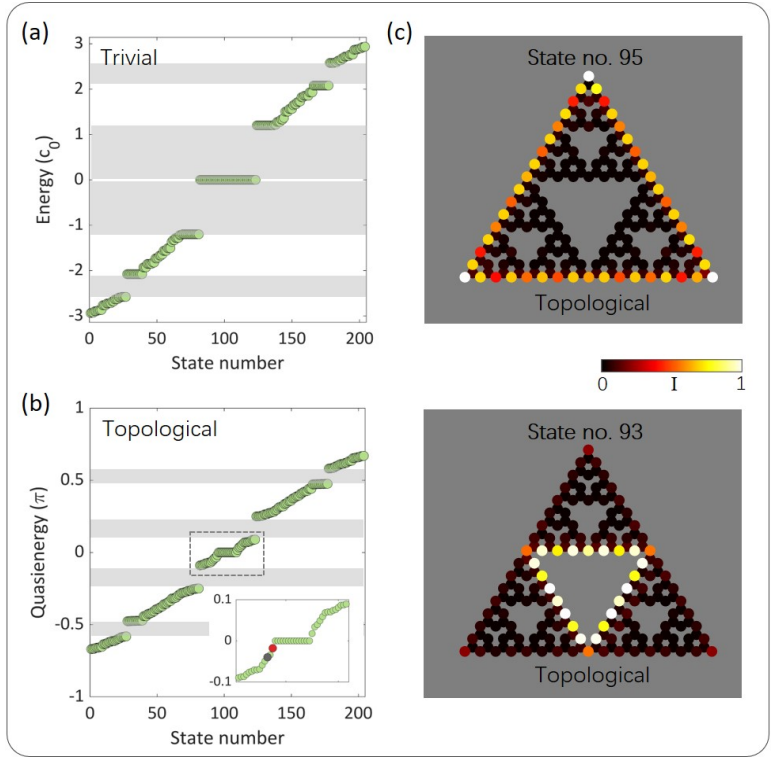

Figure 2.

Eigen states of the fractal lattice. (a) Energy spectrum with $A(z) = 0$ (straight waveguides). The spectrum of this non-topological system displays a large central gap (large gray region), with a flat band in mid-gap made up of immobile degenerate states. (b) Quasienergy spectrum with $A(z) \neq 0$ and $A_0 = kR\Omega$. The inset shows the enlarged view of the center box. The shaded regions mark quasi-gaps: regions within which there are no eigenstates. In this topological fractal system, the edge states from either side of the flat-band evolve into non-degenerate unidirectional edge states. (c) Field intensity patterns of two eigenstates localized at external and internal edges of the fractal lattice (states number 93 and 95, grey and red dots, respectively). The color bar indicates the intensity (normalized to the peak intensity in each state). The parameters used are: ambient refractive index $n_0 = 1.45$, coupling strength $c_0 = 1.9 cm^{-1}$, wavelength $\lambda = 0.633 \mu m$, helix radius $R = 10 \mu m$, longitudinal frequency of the helix $\Omega = 2\pi cm^{-1}$, lattice constant $a = 14\sqrt{3} \mu m$.

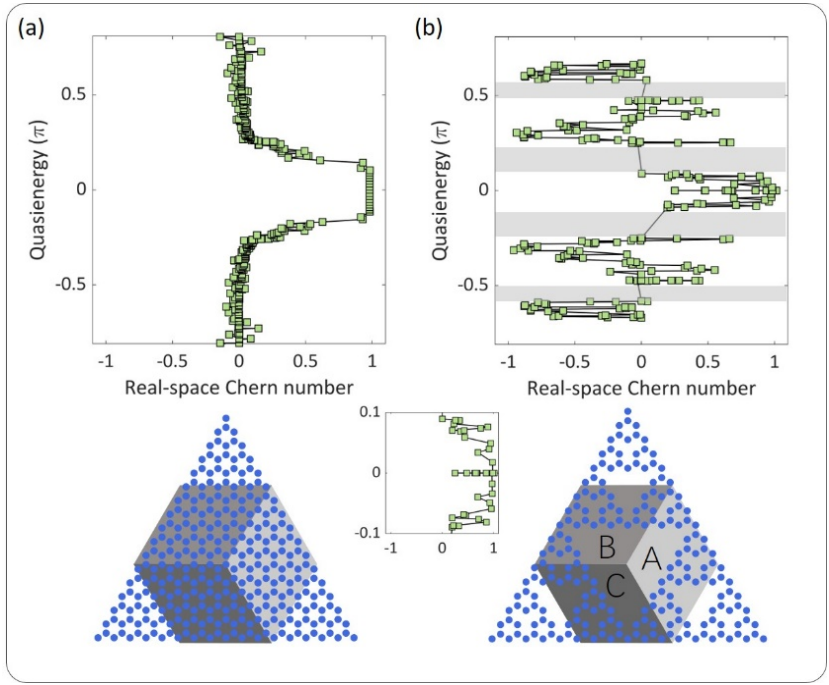

Figure 3.

Real-space Chern number as a function of quasienergy for the fractal (a) and honeycomb (b) lattices with the same nonzero artificial gauge field. The lower panels present the fractal and honeycomb lattices, which are both triangular-shaped. When calculating the real-space Chern number, the hexagons are divided into three regions with different shades of grey, each enclosing many waveguides. The center inset shows the enlarged view of the quasienergy range -0.1-0.1. The parameters of the lattices are the same as in Fig. 2.

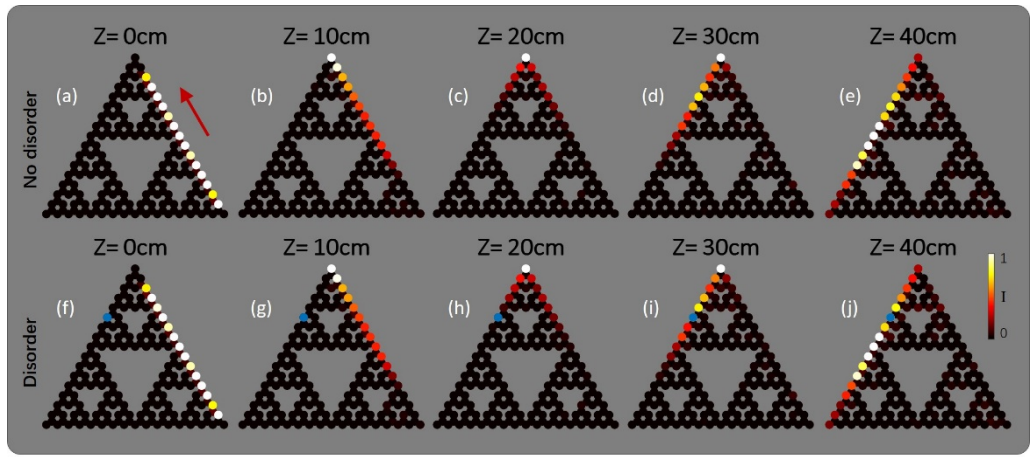

Figure 4.

Tight-binding simulations of an edge wavepacket propagating in a fractal lattice. (a-e) Evolution of topological edge states in the fractal SG(4) lattice. (a) Intensity distribution of the initial field constructed from a superposition of truncated topological edge states in the fractal lattice. (b-e) Intensity distribution at propagation distances $Z = 0, 10, 20, 30, 40$cm. (f-j) Evolution in the fractal lattice containing on-site disorder of $0.1c_0$, whose position is marked by the blue dot. The wavepacket displays topologically-protected edge transport around the corners and is unaffected by the disorder. The color bar indicates the field intensity. The parameters for the numerical simulation are the same as in Fig. 2.

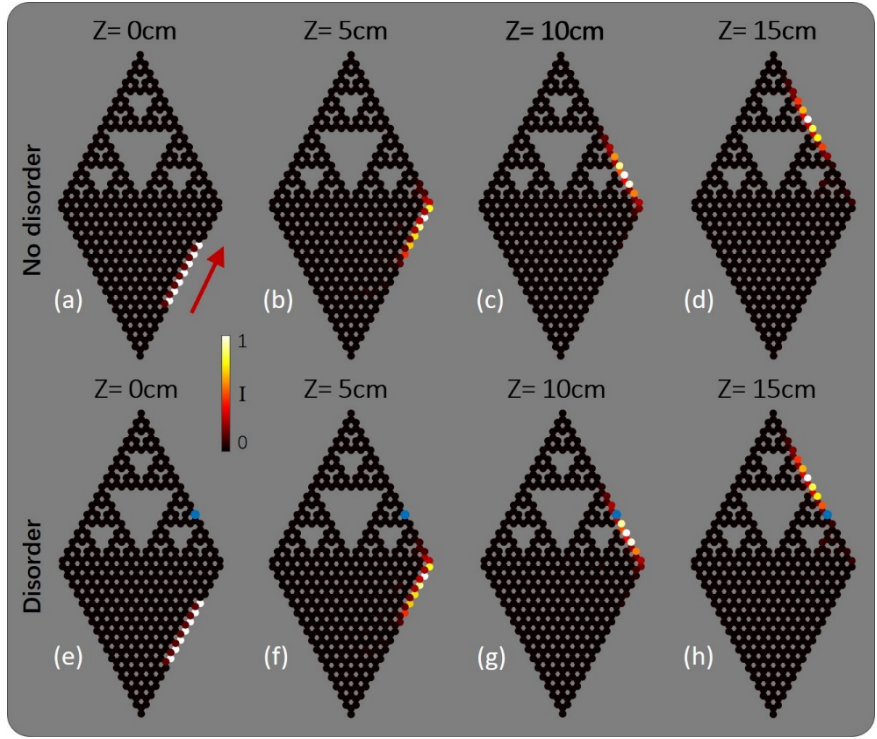

Figure 5.

Tight-binding simulations of an edge wavepacket propagating in a hybrid lattice consisting of the fractal and honeycomb lattices. The initial wavepacket (a,e) is constructed from a truncated edge state of the honeycomb lattice. (a-d) Propagation from the honeycomb region into the fractal region, displayed at propagation distances $Z = 0, 5, 10, 15$cm. (f-h) Propagation into the fractal region containing a defect (blue dot) in the form of on-site disorder of $0.1c_0$. The wavepacket exhibits propagation along the edges, around the corners, and bypassing disorder. The color bar indicates the field intensity. The parameters for numerical simulations are the same as in Fig. 2.